**Mobility of Lactose in Milk Powders**

Armin Afrough,[1] Pernille Andersen,[1] Tanja Ninette Angelika Weihrauch,[1] Dennis Wilkens Juhl,[1] Serafim Bakalis,[2] Kirsten Gade Malmos,[3] and Thomas Vosegaard[1]

[1]*Department of Chemistry and Interdisciplinary Nanoscience Center, Aarhus University, Denmark, [2]Department of Food Science, University of Copenhagen, Denmark, [3]Arla Foods amba, Denmark*

**Abstract**

Lactose is the major component of milk powders and is normally found to be in a glassy/amorphous state. During storage, lactose is known to participate in physicochemical processes, including crystallization on the surface and reaction with proteins such as β-lactoglobulin. Lactose needs to be mobile to participate in such processes. However, there is a lack of evidence of its mobility in milk powders. In this study, we demonstrate that some of the lactose becomes mobile when milk powders are exposed to humid air – an inappropriate storage condition. This mobility is evidenced by peaks in magic angle spinning $^1$H NMR spectra of milk powders in the range of 3.5 ppm to 4.0 ppm, which stem from lactose molecules displaying considerable rotational mobility. These signals have a longitudinal relaxation time constant $T_1$ similar to that of mobile water according to 2D $T_1$-$\delta_{1H}$ experiments under magic angle spinning. Furthermore, 2D $^1$H-$^{13}$C HSQC magic angle spinning experiments of skim milk powder demonstrate the same fingerprint as that of lactose in the solution, confirming our observations.

**Keywords**

Milk powder, nuclear magnetic resonance (NMR), magic angle spinning (MAS), lactose, $^1$H NMR, mobility, storage stability

**Introduction**

Milk powders are complex multicomponent systems comprising lactose, proteins, minerals, and fat – with lactose being a major component [1]. In spray-dried milk powders, lactose is in glassy/amorphous state [2], which influences several functional properties. For example, during storage – particularly under exposure to humid air or high temperatures – lactose may become mobile and participates in a variety of physicochemical processes that alter product quality [3–5]. One such process is the crystallization of lactose on the milk powder surface, where scanning electron microscopy (SEM) has shown the formation of crystals [6]. Lactose can also participate in chemical reactions with proteins such as β-lactoglobulin, $\alpha$-lactalbumin, and $\alpha_{S2}$-casein via their lysine amino acid groups [3,4,7–10]. Furthermore, lactose degradation products can promote protein aggregation in milk powders during storage [11,12].

It is difficult to measure changes in the content and state of the different constituents of milk powders as they undergo different processing and storage conditions. Techniques such as infrared and Raman spectroscopy are well-established for assessing the composition or structural properties of milk powders [13] and they are indeed sensitive to molecular motion [14]. However, although Raman – and to some extent infrared – spectroscopy could distinguish between different lactose forms, their routinely applied measurement modes generally provide only comparatively limited insight into the wider molecular-state changes that develop under varying processing and storage conditions [15]. Nuclear Magnetic Resonance (NMR) is a spectroscopic technique that is inherently and explicitly sensitive to the (rotational) mobility of molecules. This ability makes NMR one of the key methods to investigate the state of water in food drying [16].

$^1$H NMR is a method that is normally used to probe the chemical environments in a sample, e.g. by identifying different metabolites or chemical groups. In liquids, the Brownian motion of the molecules ensures the averaging of the anisotropic nuclear spin interactions (e.g. $^1$H-$^1$H homonuclear dipole-dipole couplings) leading to narrow lines. In solids, these interactions are present, and lines become very broad. Sample rotation – the so-called magic-angle spinning (MAS) – reduces the line widths [17,18], but the achieved line narrowing depends significantly on the state of the sample, i.e. whether the sample is solid (amorphous or crystalline) or mobile (gel-like). This makes $^1$H MAS NMR a very good probe to determine the mobility of species in the sample.

It is hypothesized that the mobility of lactose and other reactants enable participation in physicochemical processes [19] [19], although the direct evidence of lactose mobility in milk powders is lacking. Clarifying whether, and to what extent, lactose molecules become mobile under different storage conditions is therefore important to understanding how they participate in crystallization and chemical reactions in milk powders – affecting the powder stability and thereby shelf life. We further hypothesize that it would be possible to observe the mobility of lactose in milk powders using NMR spectroscopy, here investigated for milk powders exposed to humid air [17,18,20,21].

The objective of this work is to investigate the state of lactose in skim and fat-filled milk powders exposed to humid air. Fat-filled milk powder is an affordable alternative to whole milk powder, in which vegetable fat replaces the natural milk fat. We chose these two types of milk powders to investigate both a well-studied system (skim milk powder) in addition to one for which relatively few investigations exist (fat-filled milk powder [2,22]. For the first time, we demonstrate the rotational mobility of lactose molecules in milk powders after exposure to humid air. $^1$H MAS

NMR indicates that lactose exhibits a rotational mobility that is consistent with a liquid-like behavior. This mobile lactose can participate in physicochemical changes in the milk powder exposed to high humidity levels as extensively reported in the literature [23]. The direct NMR observation of rotational mobility of lactose provides molecular-level support to the humidity-driven quality issues in the storage of milk powders, such as caking and Maillard reactions, and reinforces the need for strict moisture control during storage.

**Results and discussion**

To investigate whether lactose becomes mobile in milk powders under humid storage conditions, we examined the surface morphology of milk powders by scanning electron microscopy. Milk powders stored under dry conditions have a smooth surface, as shown in Figure 1a. After exposure to humid air, however, lactose crystals emerge on the surface of milk powder particles [2,3,6,24] – as shown in Figure 1b-c. The appearance of new lactose crystals on the surface of milk powders implies lactose mobility. Lactose must acquire sufficient mobility to form crystals at their surface. In the following, we demonstrate through $^1$H MAS NMR experiments that mobile lactose molecules are indeed observable, providing new insight into how lactose may participate in crystallization, and reaction with proteins in milk powders.

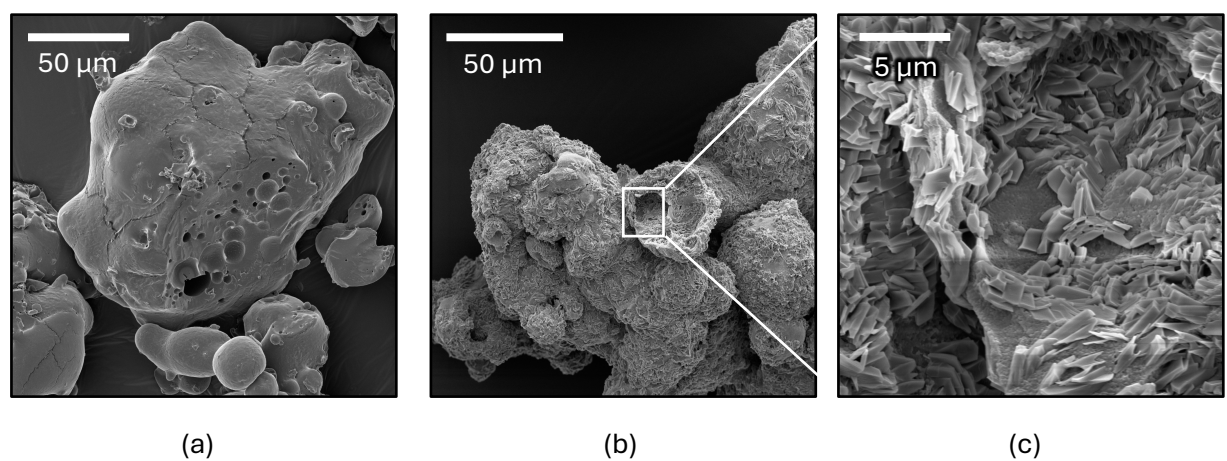

(a)  (b)  (c)

Figure 1 – Lactose crystallization on the surface of milk powder particles with exposure to humid air. Scanning electron microscopy images of skim milk powder is shown before exposure to humid air (a) and after an exposure-dehydration cycle with 75% relative humidity at room temperature for three days (b). The small lactose crystals in the frame of (b) are shown in subfigure (c). Images are acquired in the secondary electron mode with an acceleration voltage of 5 kV. Scale bars are 50 µm in (a,b) and 5 µm in (c).

NMR experiments were performed on both fat-filled and skim milk powders. Figure 2 shows the static and MAS $^1$H NMR spectra of fat-filled and skim milk powders exposed to air at different relative humidities. The static spectra of Figures 2a-b reveal two main peaks at 1.29 ppm and 4.8 ppm, which are attributed to fat and water, respectively. As expected, the water peak intensity increases with relative humidity, while the fat peak remains largely unchanged. No indication of lactose is seen in Figures 2a-b. By spinning the samples at moderate rates of 2.5 kHz to 5 kHz, the resolution is significantly improved as shown in Figure 2c-d, as mobile components of the milk powder display sharper lines (e.g. fat), while rigid components (e.g. proteins and crystalline/amorphous lactose) remains very broad. The overall result of MAS at these moderate rates is that the spectra reveal much more details than the static spectrum.

Figures 2c-d report the peak assignment of the ¹H MAS spectrum. Water has a chemical shift of 4.9 ppm. Peaks corresponding to functional groups of vegetable and milk fat in fat-filled and skim milk powders, respectively, are marked in numbers with an assignment guide on Figure 2e. Vegetable fat in fat-filled milk powder is a mixture of triglycerides with an average chain length of 16.1 ± 1.6 (determined by the relative contribution of fat peaks). The triglycerides display aliphatic protons in fatty acids (signals 1-6), glycerol backbone hydrogens (peaks 7 and 8), and hydrogens on double-bonded carbons associated with unsaturation on fatty acids (signal 9) [25–27].

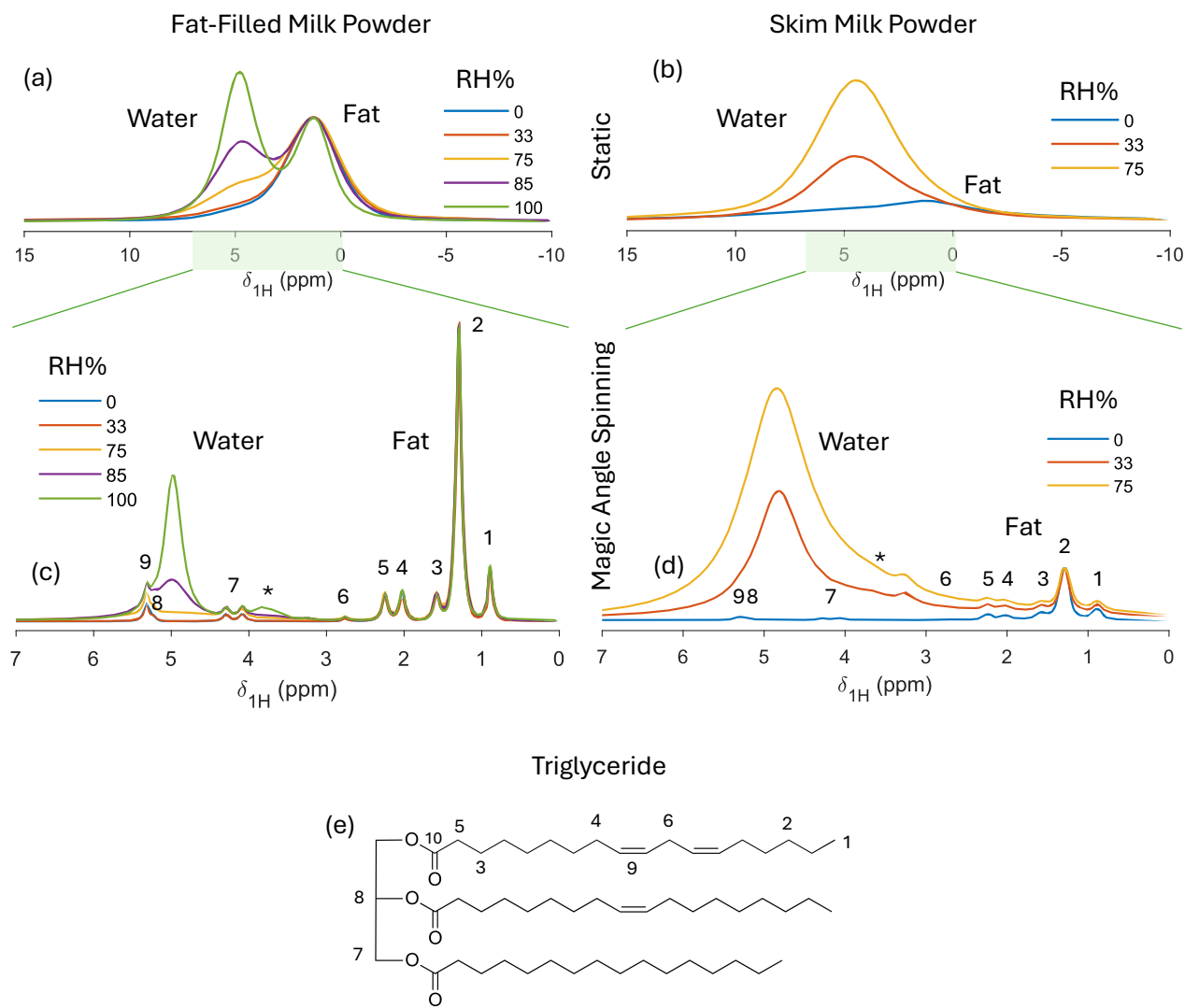

Figure 2 – ¹H NMR spectra of fat-filled (a, c) and skim (b, d) milk powders following exposure to humid air at varying levels of relative humidity (RH%) at room temperature. Panels (a, b) are static spectra and only show broad peaks corresponding to water and fat. Magic angle spinning significantly reduces the linewidth of the peaks (c, d) and even makes it possible to observe peaks corresponding to different functional groups of triglycerides in vegetable and milk fat (e): 1 to 6 for aliphatic from fatty acids, 7 to 8 aliphatic from glycerol backbone, and 9 unsaturation on the fatty acids – see the assignment guide on the molecular structure. Mobilized lactose is marked with * in the magic angle spinning spectra and is clearly visible at high hydration levels – while it is not visible in static ¹H spectra (a).

While most peaks visible in the MAS spectra could be assigned to the different hydrogens in fat, we observed a set of peaks in the $^1$H chemical shift range of 3.5 ppm to 4.0 ppm in fat-filled and skim milk powders after exposure to humid air. The chemical shifts of these peaks matches the resonances of lactose in aqueous solution (see Figure SM2a). Hence, we attribute these peaks to a fraction of lactose that has gained significant rotational mobility – almost similar to that of a liquid – when milk powders have absorbed water. These peaks are particularly pronounced at higher humidities. In the rest of this section, we provide our rational and experimental evidence for this assignment. In particular, we differentiate the observed mobile lactose from solid lactose that will not appear visible at such low MAS rates, due to their broad peaks.

In liquids, the rotational motion of molecules averages the nuclear spin interaction of nuclei such that the anisotropy of the nuclear interactions is averaged away. As a rule of thumb, this averaging occurs if the rotational correlation time is shorter than the characteristic time of the interaction. For $^1$H-$^1$H dipole-dipole interactions, which are up to approximately 20 kHz, the characteristic time would be 1/20 kHz = 50 µs. However, as the homonuclear ($^1$H-$^1$H) dipole-dipole interactions are homogeneous [28], it requires significantly shorter (but difficult to quantify) rotational correlation times to effectuate this averaging.. To illustrate the line narrowing achieved by molecular motion, Figure 3 shows the $^1$H MAS NMR spectrum of milk powder exposed to humid air (a) along with spectra of lactose in solution (b) and in solid (crystalline) form (c). Indeed, the spectrum of dissolved lactose shows a set of very narrow peaks (~ 1 Hz full-width-at-half-height (FWHH) linewidth at 950 MHz $^1$H frequency), while the spectrum of solid lactose shows a broad hump, where individual peaks cannot be identified.

In this context, it is important to note that the total intensity of the spectrum – the area below the curve – is proportional to the total number of hydrogens in the sample. When signals appear as narrow peaks, these peaks are much higher. Therefore, the vertical scales in Fig. 3b and 3c differ by several orders of magnitude. For samples containing both dissolved and solid lactose, the signals from solid lactose may be too weak to be observed – seemingly absent.

Chemical shifts in solids may differ somewhat from those in liquids due to crystal packing, and dipole-dipole couplings may lead to differential line broadening. It is however a reasonable approximation to model spectra of lactose with reduced mobility by applying uniform line broadening to all peaks – despite differences between solid and liquid chemical shift. Figure 3d shows an expanded $^1$H NMR spectrum of milk powder exposed to humid air alongside the spectrum of dissolved lactose with an additional line broadening of 60 Hz (FWHH). Indeed, the unassigned peaks in the milk-powder spectrum match well with the lactose signals. Similarly, Figure 3e shows the spectrum of solid lactose alongside the spectrum of dissolved lactose, this time with a line broadening of 1000 Hz (FWHH).

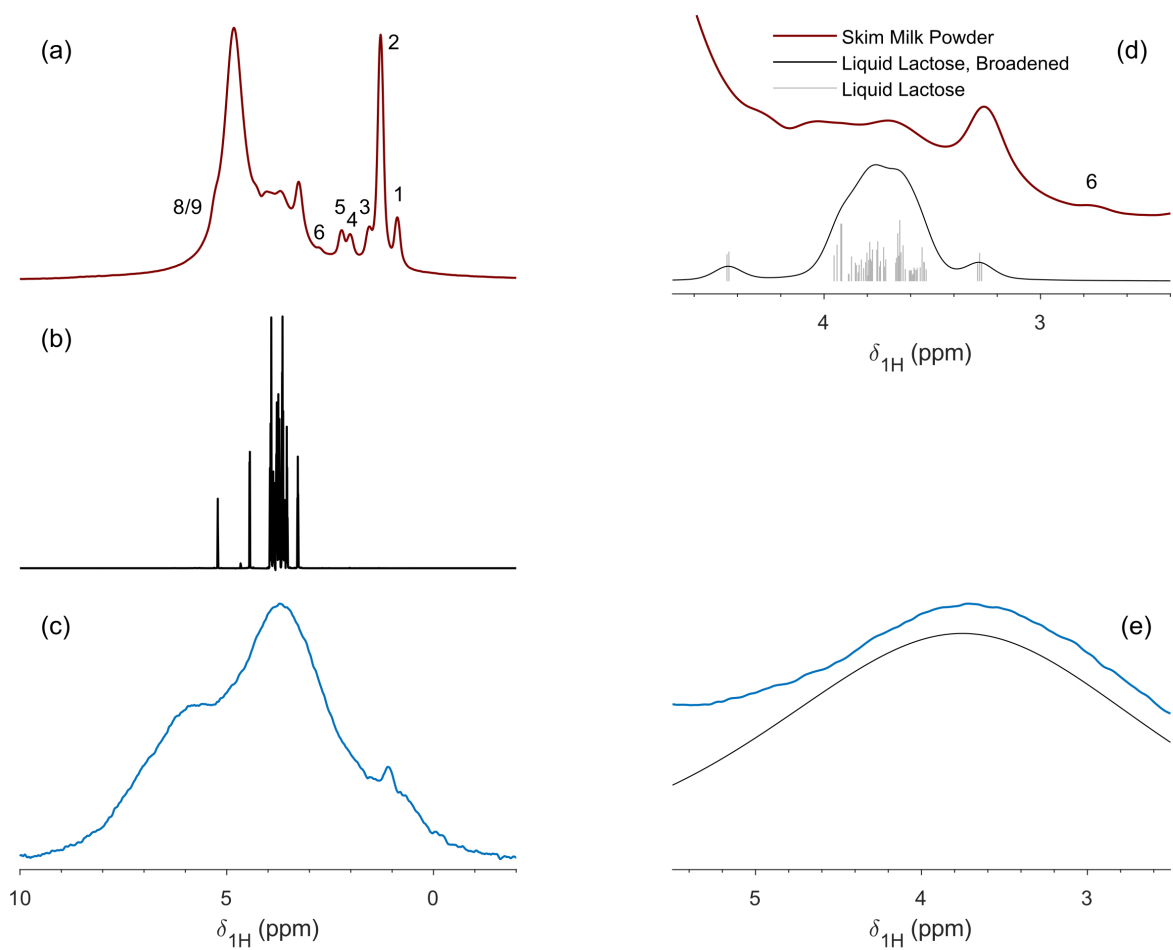

Figure 3. $^1$H NMR spectra of (a) skim milk powder exposed to 75% relative humidity (with a fraction of ~ 95% $D_2O$) for one day, (b) dissolved lactose, and (c) solid lactose under MAS. Expansions of the spectra showing (d) the spectrum of milk powder along with the spectrum of dissolved lactose with a line broadening of 60 Hz, and (e) the spectrum of solid lactose (blue) with the spectrum of dissolved lactose with a line broadening of 1000 Hz. See Figure SM5 for the same Figure produced for fat-filled milk powder exposed to 100% relative humidity.

It is important to note that these observations imply the rotational mobility of lactose – in contrast with the translational motion. NMR has the ability to probe both rotational motion [29] and translational motion [30] of chemical species. Translational motion is measured by diffusion NMR experiments, but such experiments require relatively fast translational diffusion due to the nature of the experiments, and the confined environment of milk powders results in too slow translational diffusion for normal diffusion NMR experiments, but this will be a topic of future investigations.

MAS spectra of milk powders may be fitted with peaks representing water, fat, and lactose. Figure SM1 (Supplementary Materials) shows such a fit where all functional groups are fitted simultaneously. The 3.5 ppm to 4.0 ppm region in $^1$H spectra of milk powders exposed to humid air is seen as a constellation of overlapping peaks. By applying line broadening and scaling (two parameters for lactose), we fitted the solution spectrum of lactose to the MAS spectra. Relative to lactose in solution, the broader linewidths of lactose peaks in powders exposed to humid air indicate that in powders, although lactose is still appreciably mobile, its rotational motion is slower than that of freely dissolved lactose. By fitting these spectra as a function of the relative humidity, we acquired the relative intensity of water and lactose at different relative humidities with respect to the intensity of the -CH$_3$ peak of triglycerides, as shown in Figure 4 for fat-filled milk powder. The intensity of mobile lactose in skim milk powder was comparable to that of fat-filled milk powder (results not shown). The intensity of mobile lactose shows a correlation with the water content.

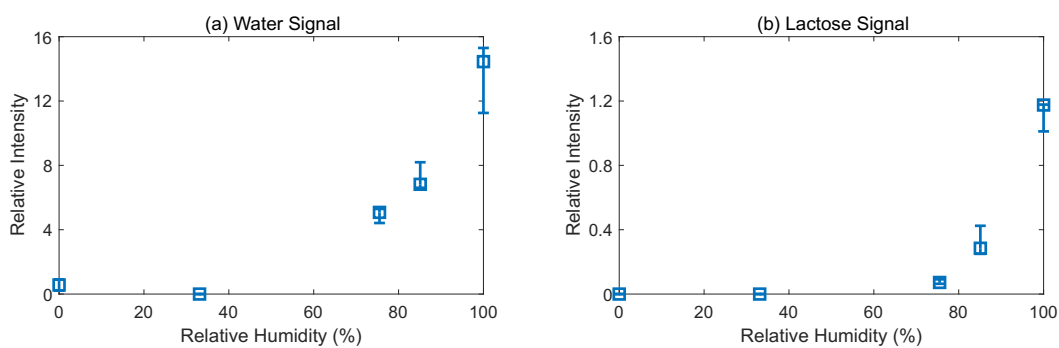

Figure 4 – The intensity of (a) water and (b) lactose peaks for fat-filled milk powder as a function of relative humidity. The intensity is relative to the -CH$_3$ peak in triglycerides. Mobile lactose content is correlated with the water signal observed in $^1$H MAS spectra.

Although the relationship between the mobility of lactose with the existence of water was observed as a correlation in Figures 2-4, it can also be derived phenomenologically. The 2D longitudinal relaxation $T_1 - \delta\ {}^1_H$ correlation map measures a $T_1$ relaxation time distribution for each binned chemical shift value providing information on the dynamics and environment of molecules. Figure 5 shows such a $T_1 - \delta\ {}^1_H$ correlation map where lactose signals are observed as a band in the 3.5 to 4.0 ppm chemical shift range at $T_1$ = 2.3 s. This band for lactose is close to that of water from $\delta_{1H} = 4$ to 6 ppm at $T_1$ = 2.0 s (62% of total water signal). The closeness between these two longitudinal relaxation parameters is an indication of cross-relaxation and/or exchange processes between lactose and water in milk powders [31]. We assign the water peak at $T_1 = 0.68$ s (22%) and $T_1 = 0.22$ s (12%) to water associated to micellar casein and whey proteins, respectively. The intensity of these peaks is consistent with the simulated moisture sorption of skim milk powder at 20°C: 60% by lactose, 32% by micellar caseins, and 8.5% by whey proteins – with a total of 9.9 g water per 100 g dry milk powder and disregarding the effect of ash [32].

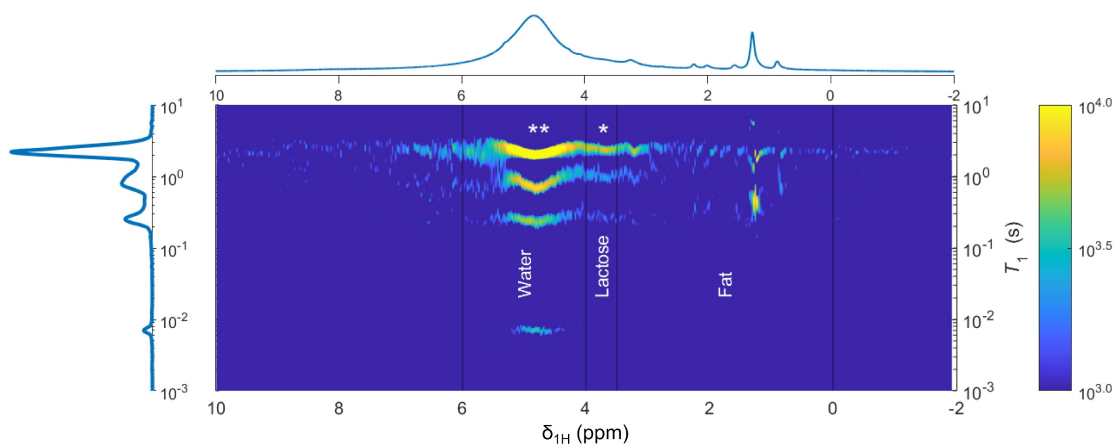

Figure 5 – NMR $^1$H longitudinal relaxation time $T_1$ – chemical shift $\delta_{1H}$ correlation map of skim milk powder exposed to air with 53% relative humidity (80% D$_2$O, 20% H$_2$O) for three days. $^1$H in water dominates the NMR signal with four peaks in the $T_1$ distribution from $\delta_{1H} = 4$ to 6 ppm. The most intense water peak (**) at $T_1$ = 2.0 s is for water associated with lactose (*) shown as a band from $\delta_{1H} = 3.5$ to 4 ppm and $T_1$ = 2.3 s. MAS averages magnetic susceptibility differences between water and other components of milk powder particles (casein and whey proteins, lactose, oil). Proteins are not visible in the spectra. NMR data was acquired at 400 MHz $^1$H frequency with 5 kHz magic angle spinning. The color scale is logarithmic to highlight small signals.

In addition to lactose, there are significant protein quantities in milk powders that may be mobilized with the absorption of water in the solid matrix of milk powders. The mobility of $^1$H sites in casein and, in particular, whey proteins should be ruled out. We acquired solution-state $^1$H NMR spectra of major components of milk powders – whey proteins, casein proteins, and lactose – in D$_2$O with 1D $^1$H and 2D $^1$H -$^{13}$C HSQC spectra. These spectra are reported in the Supplementary Materials (see Figures SM2-3). Solutions of casein and whey protein powders in D$_2$O produced a constellation of peaks in the 0.5 – 3.5 ppm and 6 – 8 ppm regions of the spectra (see Supplementary Materials) that were not observed in the MAS spectra of milk powders exposed to humid air. The 1D $^1$H spectra as well as 2D HSQC spectra of lactose determined that the source of signal in the 3.5 ppm to 4.0 ppm range is indeed lactose.

An excellent match to the HSQC data of lactose was found for the MAS spectra of milk powders exposed to humid air. Two-dimensional HSQC experiments correlate the chemical shift of $^1$H nuclei neighboring with $^{13}$C nuclei. HSQC spectra are seen as a constellation of peaks on a 2D plot where the center of each peak is for a $^1$H-$^{13}$C neighboring nuclear pair in a chemical structure. The HSQC data of lactose dissolved in D$_2$O is shown in Figure 6a. The elevated water content increased signal intensity in the 3.5 ppm to 4.0 ppm such that it could be measured by a 2D method overnight. The similarity between the HSQC correlation plots in the milk powder versus that of the liquid lactose demonstrate that the signal in the 3.5 ppm to 4.0 ppm is indeed dominated by lactose.

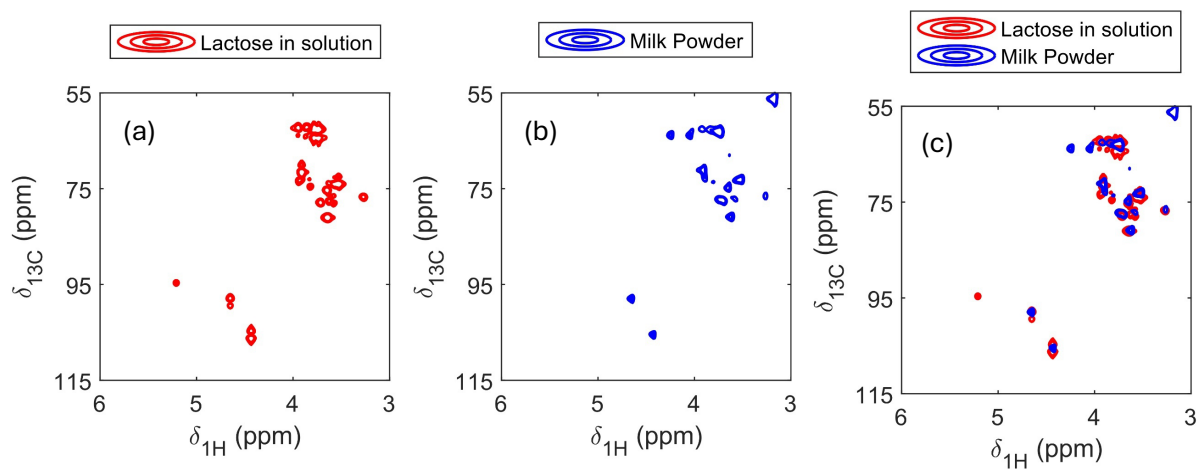

Figure 6 – $^1$H-$^{13}$C HSQC chemical shift correlation spectra of (a) lactose in aqueous solution, (b) skim milk powder exposed to humid air (100% relative humidity for one day), and (c) their superposition. The constellation of peaks in milk powder exposed to humid air exactly matches the fingerprint of lactose HSQC $^1$H-$^{13}$C spectra. This match is clearly shown in (c) by superimposing a contour of the lactose peaks (red) on the that of the milk powder (blue). The two-dimensional spectrum in (a) was measured at $^1$H frequency of 950 MHz, and that of (b) with high-resolution magic angle spinning at 700 MHz. The similarity between the two patterns demonstrates that the signal observed in the 3.5 ppm to 4.0 ppm spectra previously observed is indeed lactose. Peaks other than lactose observed in (b) are related to milk fat. The spectra are referenced to DSS.

For the first time, with evidence from solid-state NMR spectroscopy, we were able to show the increased mobility of some of lactose in milk powders. In molecular simulations, it has been shown that water has a catalytic effect on the ring opening of glucose [33]. For glucose, the most time-consuming steps of the ring-opening reaction are the orientational rearrangements of water molecule(s) participating in the proton transfer(s) and the final extension of the newly-formed aldehyde chain [33]. We hypothesize that similar processes could be relevant to lactose in milk powders and hence reducing the mobility of water and lactose molecules can reduce their participation in physicochemical processes. Further studies along these lines will be pursued.

**Materials and Methods**

*Samples*

Solid- and liquid-state NMR experiments were performed on commercial bovine dairy powders: fat-filled milk powder, skim milk powder, whey protein concentrate, micellar casein isolate, and lactose all from Arla Foods amba (Viby, Denmark). Fat-filled milk powder had a mass-based composition of 41% lactose, 22% proteins, 28% fat (palm oil), and 5% sucrose in addition to minerals. Skim milk powder had a mass-based composition of 54% lactose, 32% proteins, and 0.5-0.9 % fat (milk fat), in addition to 9 % minerals. Whey protein concentrate is composed of 76-80% protein and contains maximum 9% lactose, 10% fat, 3.5% ash and 6% moisture.

Micellar casein isolate is composed of 86% by mass protein where 97% is casein micelles; it also contains maximum 2% lactose, 1% fat, 8% ash, and 5% moisture. Lactose monohydrate was derived from cheese whey by ultrafiltration and crystallization and constitutes at least 99% of the lactose in lactose powder with maximum 0.3% sulphated ash, 0.2% moisture and 0.2% protein, and 5.2% total water.

Magic angle spinning experiments were undertaken on either dry powders, or powders exposed to humid air in a desiccator. The relative humidity of air in the desiccator was controlled by saturated salt solutions at room temperatures from 20°C to 25°C. Salts used include: $MgCl_2$ at 33%, NaCl at 75%, and KCl at 85% relative humidity. Not all salt solutions were used for all milk powder types. For fat-filled milk powders, only $H_2O$ was employed in preparing salt solutions. For skim milk powders, a mixture of $H_2O$ and $D_2O$ were used. We inspected some samples by a stereomicroscope before and after MAS to ensure that the structure of the samples is not compromised.

For liquid-state NMR experiments, 1D $^1H$ FID with water suppression [34] and 2D $^1H$-$^{13}C$ HSQC [35] methods measured solution samples of lactose and milk protein powders, whey and caseins, dissolved in $D_2O$. High-resolution magic angle spinning HSQC experiments were conducted on skim milk powder exposed to a relative humidity of 100% for one day. Milk powders were exposed to varying levels of relative humidity between one to three days, except for 100% relative humidity where exposure was for one day only (to avoid the growth of mold). All liquid- and solid-state spectra were referenced to either a solution of sodium trimethylsilylpropanesulfonate (DSS) in $D_2O$ (for the majority of the samples), or solid adamantane.

Selected samples were investigated with scanning electron microscopy before and after exposure to humid air (75% for three days). Samples were mounted on aluminum stubs with mounted two-sided carbon tapes. Milk powders were placed on the carbon tape. Compressed air removed milk powder particles that were not stuck to the carbon tape. The samples were vacuumed and coated with 10 nm platinum. Imaging was performed with the secondary electron mode with a Clara S8151 instrument from TESCAN XRE operating at 5 kV.

*NMR Experiments*

NMR spectra were acquired with three spectrometers. Liquid-state spectra were acquired with a Bruker Avance III HD 950 MHz spectrometer using a 5 mm TCI liquid state cryoprobe. HR-MAS HSQC spectra of skim milk powder, and MAS spectra of solid lactose were acquired with a Bruker Avance III 700 MHz spectrometer: HSQC spectra using a 4-mm HR-MAS probe spinning at 4 kHz, and solid-state NMR spectra using a 1.3-mm $^1H$-$^{13}C$-$^{15}N$-$^2H$ MAS probe spinning at 60 kHz. Solid-state NMR spectra of milk powders were acquired with a Bruker 400 MHz spectrometer using a 4-mm $^1H$-X-Y MAS probe at spinning speeds between 2.5 kHz to 5 kHz. Spectra were processed in EasyNMR, MATLAB, and TopSpin. All data is available in Zenodo (https://doi.org/10.5281/zenodo.15979671). The spectra were acquired at a sample temperature of 25°C by setting the temperature unit to calibrated set points for static and MAS experiments [36]. The repetition delay for milk powders was set to between 15 s to 40 s to exceed five times the relaxation time of water and fat in milk powders. Two-dimensional $T_1 - \delta\ _{^1H}$ measurements were undertaken with an inversion-recovery pulse sequence at 400 MHz with 64 delay times log-normally distributed from 10 µs to 40 s. After Fourier transformation of data in the chemical shift dimension, a multi-exponential recovery method [37] was employed to invert time-domain data of binned chemical shifts to the $T_1$ distribution.

**Acknowledgements**

Authors would like to thank Anders Bodholt Nielsen, Niels Christian Nielsen, Fábio Mariz Maia Neto, Ramon Liebrechts, Rebekka Klemmt for discussions and help with measurements. The use of NMR facilities at the Danish Center for Ultrahigh-Field NMR Spectroscopy are funded by the Danish Ministry of Higher Education and Science (Grant No. AU-1198 2010-612-181) and the Novo Nordisk Foundation (Grant No. NNF220C0075797). We are grateful for financial support from the EU H2020 project PANACEA (Grant No. 101008500), the EU Horizon-Europe project r-NMR (Grant No. 101058595), Arla Foods amba and Innovation Fund Denmark (1063-00031B). The Carlsberg Foundation (Grant no: CF20-0364) and iMAT are acknowledged for funding the TESCAN Clara SEM used in this work.

**Author Declarations**

**Conflict of Interest**

The authors have no conflicts to disclose.

**Data Availability**

The data that support this study are openly available in Zenodo at https://doi.org/10.5281/zenodo.15979671.

**Supplementary Materials**

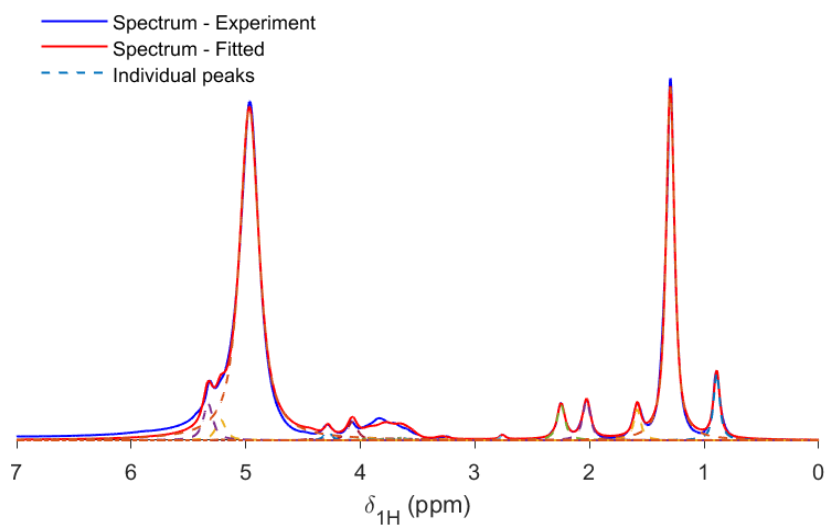

Figure SM1 – Fitting of a $^1$H MAS NMR spectrum with peaks for water, triglyceride, and lactose functional groups. The -CH$_3$ at 0.891 ppm has an integral intensity of 0.0177, in comparison to that of 0.2708 for water centered at 4.97 ppm and lactose with a total intensity of 0.0208.

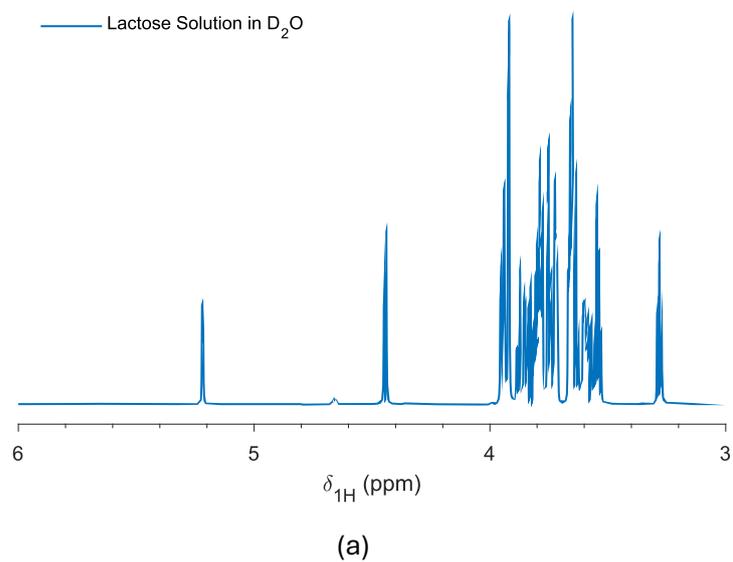

(a)

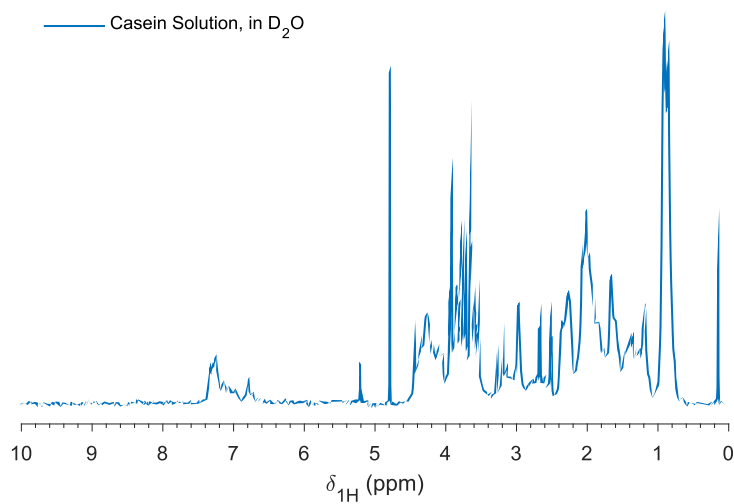

(b)

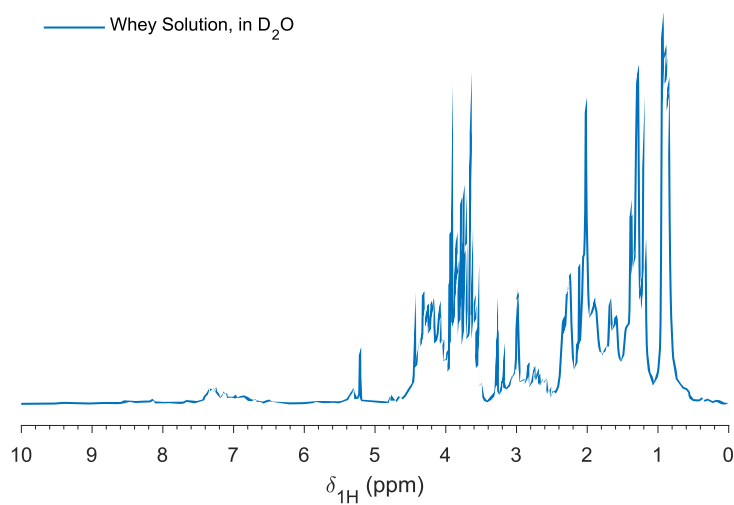

(c)

Figure SM2 – Liquids-state $^1$H NMR spectra of lactose, casein, and whey solutions at 950 MHz. Spectra are referenced to DSS. In the $^1$H NMR spectrum of lactose at 950 MHz, the peaks are between 3.2 ppm and 5.3 ppm, where most of the peaks are in the chemical shift range of 3.5 ppm to 4.0 ppm. $^1$H liquid NMR spectrum for casein and whey proteins show significant similarity that is also shown in the solid experiments. In the liquid spectra, more detail is visible than that of solid-state measurements because of the mobility. The solubility of components in D$_2$O is $1.044 \times 10^{-3}$ g/ml for caseins, $1.046 \times 10^{-1}$ g/ml for whey, and $8.968 \times 10^{-2}$ g/ml for lactose. Solubility directly affects the signal-to-noise ration observed in NMR spectra – notice the higher noise in the casein spectrum. Casein is the most insoluble component in D$_2$O with a solubility 100 times less than that of whey protein. The insolubility of casein is because of its outward hydrophobic domains.

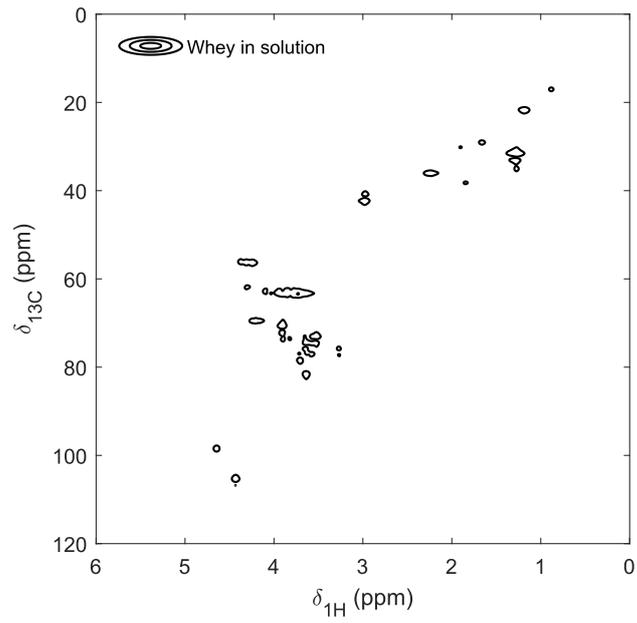

(a)

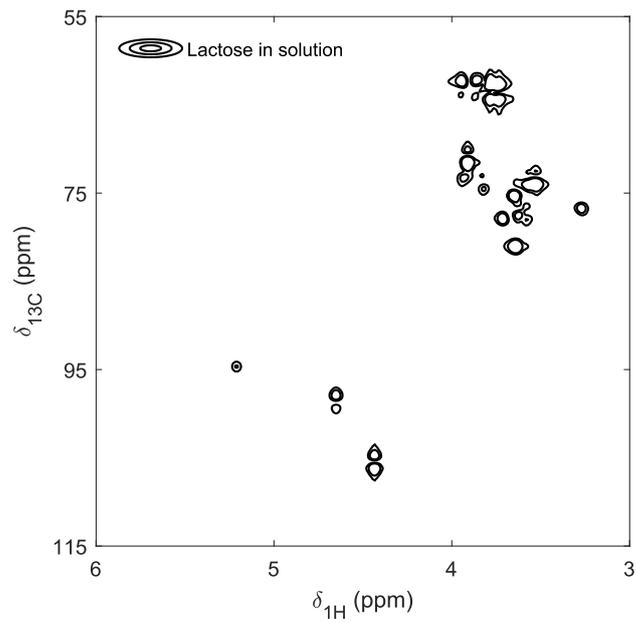

(b)

Figure SM3 – 2D $^1$H-$^{13}$C HSQC spectra of lactose and whey solutions in D$_2$O at 950 MHz. Spectra are referenced to DSS.

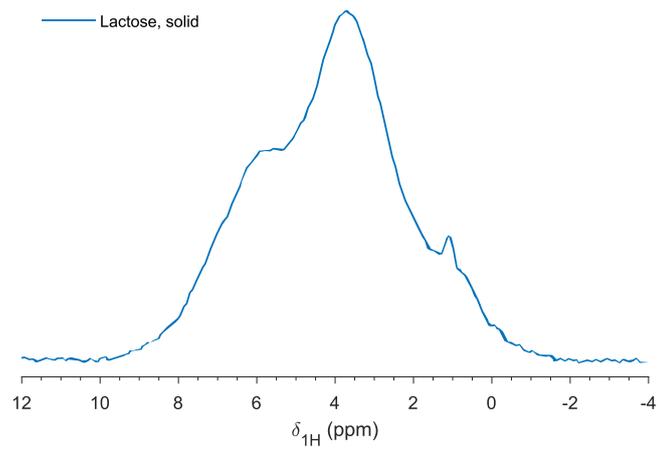

(a)

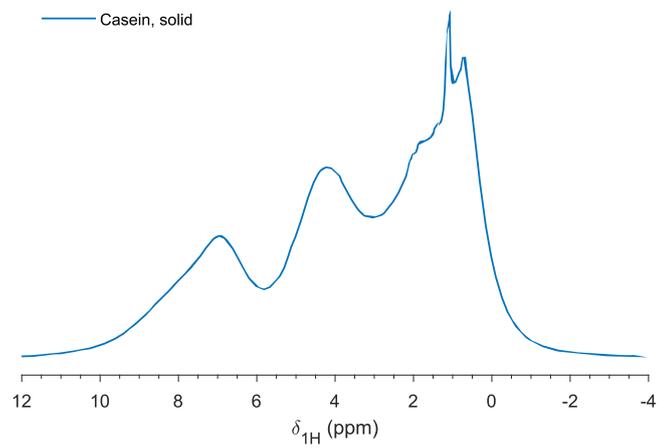

(b)

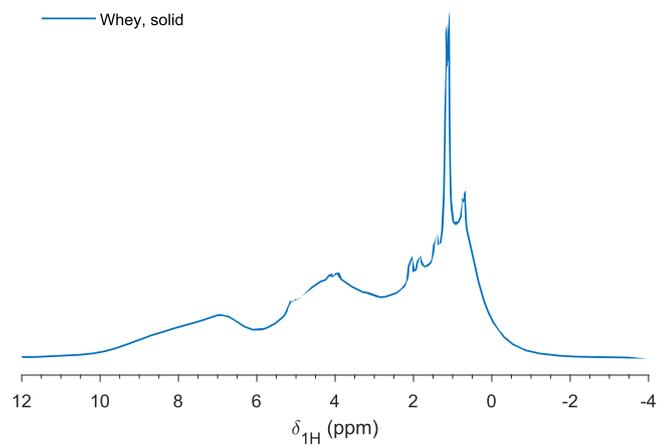

(c)

Figure SM4 – Solid-state $^1$H NMR spectra of lactose, casein, and whey powders at 700 MHz $^1$H and 60 kHz MAS.

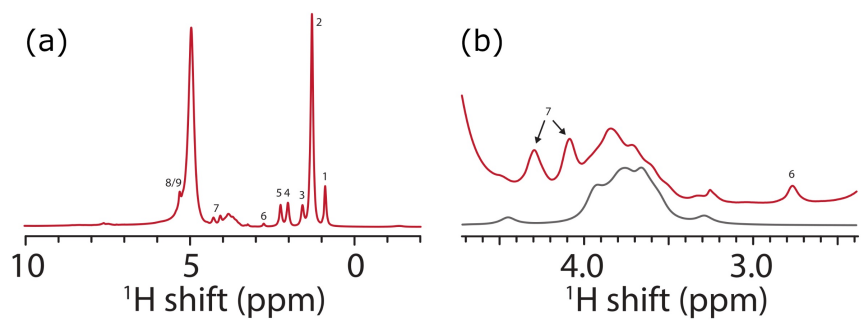

Figure SM5 – $^1$H NMR spectra of (a) fat-filled milk powder exposed to 100% relative humidity for 1 day, and (b) expansions of the spectra showing the spectrum of milk powder along with the spectrum of dissolved lactose with a line broadening of 100 Hz. See Figure 3 for the same Figure produced for skim milk powder exposed to 75% relative humidity.